\documentclass{mem}
\usepackage{natbib}\usepackage{txfonts}\usepackage{balance}
\usepackage{graphicx}
\usepackage[a4paper,breaklinks,dvipdfm]{hyperref}
\begin{document}
\def\teff{$T\rm_{eff }$}
\def\kms{$\mathrm {km s}^{-1}$}

\title{
Shocks and Cavities from Multiple Outbursts in the
Galaxy Group NGC 5813:
}

   \subtitle{A Window to AGN Feedback}

\author{
S. \,W. \,Randall\inst{1}, 
W. \,R. \,Forman, 
S. \,Giacintucci,
P. \,Nulsen, 
M. \,Sun,
C. \,Jones, 
E. \,Churazov,
L. \,David,
R. \,Kraft,
M. \,Donahue,
E. \,Blanton,
A. \,Simionescu,
\and N. \,Werner
          }


\institute{
Smithsonian Astrophysical Observatory
60 Garden St. MS-70
Cambridge, MA 02138, USA
\email{srandall@cfa.harvard.edu}
}

\authorrunning{Randall }

\titlerunning{AGN Feedback in NGC~5813}

\abstract{

We present results from {\it Chandra} observations of
NGC~5813, the dominant central galaxy in a nearby galaxy group. We
focus 
on three main results. 1) The diffuse gas shows clear signatures from
three distinct outbursts of the central AGN, with three pairs of
roughly collinear cavities. The inner two cavity pairs are associated
with unambiguous elliptical shock fronts, with Mach numbers $M \sim 1.7$ and
$M \sim 1.5$ for the inner and outer shocks, respectively. 2) The mean power
of the two most recent outbursts differs by a factor of six,
indicating that the mean jet power varies over long ($\sim 10^7$~yr) time
scales. 3) The heating from the shocks alone is sufficient to balance
radiative cooling of the gas within at least the central 10~kpc,
allowing feedback to operate isotropically at small radii. 

}
\maketitle{}

\section{Introduction}
Early {\it Chandra} and {\it XMM-Newton} observations revealed that
the amount of gas radiatively cooling to low temperatures is less than
what is predicted by the classical cooling flow models
\citep{david01, peterson01}, the implication being that the central gas must
be heated.   A 
promising heating candidate is feedback from energy injection 
by the central AGN of the cD galaxy 
\citep[for a review see][]{mcnamara07}.
Generally, studies have
found that the internal energies and ages of these cavities give
AGN kinetic powers that are sufficient, in principle, to balance
cooling \citep[e.g., ][]{birzan04}.  However, 
it is unclear how the cavity internal energy is released to heat the
ICM near the central AGN.
Outburst shocks, driven by the expansion of cavities when they are first
inflated, will also work to heat the central ICM.
Here we use {\it Chandra} observations of the galaxy group NGC~5813 to study
AGN feedback in detail.
\begin{figure}[]
\hspace{0.1in}
\resizebox{2.3in}{!}{\includegraphics[clip=true]{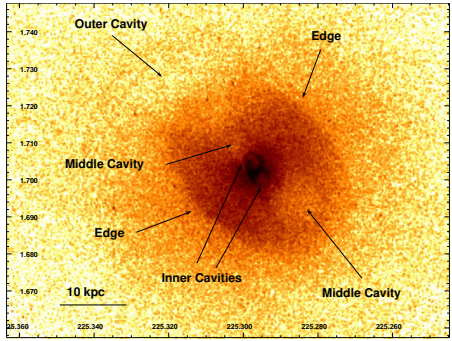}}
\caption{\footnotesize
Exposure corrected, background subtracted, 0.3 -- 2~keV {\it Chandra}
image of NGC~5813, smoothed with a 1.5'' radius Gaussian.
}
\label{ximg}
\end{figure}
\section{NGC~5813}
\begin{figure}[]
\hspace{0.2in}
\resizebox{2in}{!}{\includegraphics[clip=true]{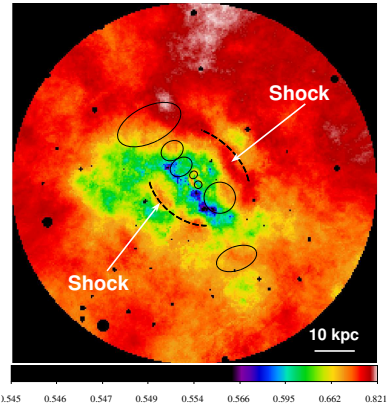}}
\caption{\footnotesize
Temperature map of NGC~5813, with the locations of the shocks and
cavities overlaid.  The color bar gives the temperature in keV.
}
\label{tmap}
\end{figure}

NGC~5813 (UGC~09655) is the central dominant member of a nearby (redshift $z =
0.006578$) subgroup in the NGC~5846 galaxy
group \citep{mahdavi05}.  It was observed for a total of
150~ks with the {\it Chandra X-ray Observatory}.  Details of the data
analysis are presented in \citep{randall11}.  The {\it Chandra} image
(Fig. \ref{ximg}) shows three pairs of collinear
cavities.
The inner two pairs are surrounded by sharp, elliptical surface
brightness edges. 
The corresponding temperature map, shown in Fig. \ref{tmap}, reveals
hot gas just inside the bright edges, indicating that these are shock
fronts.  Detailed temperature and density profiles across the edges
confirm that they are shocks, with Mach numbers $M
\sim 1.5$ and $M \sim 1.7$ for the 10~kpc and 1.5~kpc shocks,
respectively \citep{randall11}.  

\section{Discussion}

For each outburst, the total energy is
the internal energy of the cavities plus the
shock energy (the radiative energy output is negligible).
The cavity energy is estimated as $3PV$, where $P$ is the gas
pressure at the radius of the cavity. The shock energy is
estimated by matching 1D hydrodynamic point explosion
simulations with the X-ray image.
 We
find that the mean power of the current outburst
is less than that of the
previous outburst by
about a factor of six ($1.5 \times 10^{42}$~erg/s vs. $1.0 \times
10^{43}$~erg/s). 
This demonstrates that the mean jet power of the central
AGN varies significantly over the timescale between
outbursts ($\sim 10^7$~yr).

The heating of the ICM locally at each shock front can be
determined from the measured shock properties. To offset
radiative cooling in the gas, the heating mechanism is
required to increase the gas entropy $S$. 
Expressed as a fraction of the local thermal energy of the
gas $E$, the heat input at the shock front is
$(T \, \Delta S)/E = \Delta \ln (P / \rho^\gamma)$. We find this fraction
to be 10\% for the inner and 
5\% for the outer shock, respectively. Therefore,
1/0.1 = 10 (1/0.05 = 20) shocks are needed per local
cooling time of the gas to offset radiative cooling at the 1.5~kpc
(10~kpc) shock front. The cooling time of the gas
is $2 \times 10^8$~yr
($9 \times 10^8$~yr), so that the outburst interval of $10^7$~yr gives 20
(90) shocks per cooling time. Thus, the heat input from
shocks alone is sufficient to balance radiative cooling of the
gas within at least the central 10~kpc.


\bibliographystyle{aa}

\end{document}